\documentclass[useAMS,usenatbib]{mnras}

\usepackage{graphicx} 
\usepackage{times}
\usepackage{epsfig}
\usepackage{epstopdf}
\usepackage{amsfonts}
\usepackage{amsmath}
\usepackage{amsbsy}
\usepackage{bm}
\usepackage{url}
\usepackage{microtype}
\usepackage{rotating}
\usepackage{sidecap}
\usepackage{longtable}
\usepackage{lscape}
\usepackage{rotating}
 \usepackage{array,multirow,graphicx}
\usepackage[export]{adjustbox}
\usepackage{amssymb,xcolor}
\usepackage{colortbl}
\usepackage{booktabs}
\usepackage{url}
\usepackage[flushleft]{threeparttable}
\usepackage[utf8]{inputenc}
\usepackage[spanish]{babel}
\usepackage{array}
\newcolumntype{L}{>{\raggedright\arraybackslash}p{.5\linewidth}}

\def\apj{ApJ}
\def\apjs{ApJ}
\def\mnras{MNRAS}
\def\aap{A\&A}
\def\apjl{ApJ}

\def\araa{Annual Reviews}

\def\ssr{Space Science Reviews}

\usepackage[T1]{fontenc}
\usepackage{ae,aecompl}

\usepackage{newtxtext,newtxmath}

\title[XTE J1859-226: Spin and mass]{Black hole mass and spin measurements through the Relativistic Precession Model: XTE J1859+226}

\author[S. Motta et al.]{Motta, S.E.$^{1,2}$, Belloni, T.$^{1}$, Stella, L.,$^{3}$, Pappas, G.$^{4}$, Casares, J.$^{5,6}$, 
A., Mu\~noz-Darias, T.$^{5,6}$, 
\newauthor Torres, M.A.P.$^{5,6}$, Yanes-Rizo, I.V.$^{5,6}$
 \\
$^{1}$Istituto Nazionale di Astrofisica, Osservatorio Astronomico di Brera, via E.\,Bianchi 46, 23807 Merate (LC), Italy\\
$^{2}$University of Oxford, Department of Physics, Astrophysics, Denys Wilkinson Building, Keble Road, OX1 3RH, Oxford, United Kingdom\\
$^{3}$ INAF, Osservatorio Astronomico di Roma, Via Frascati 33, 00078 Monteporzio Catone, Roma, Italy\\
$^{4}$ Department of Physics, Aristotle University of Thessaloniki, Thessaloniki 54124, Greece\\
$^{5}$ Instituto de Astrof\'isica de Canarias, 38205 La Laguna, Tenerife, Spain \\
$^{6}$ Departamento de Astrof\'isica, Univ. de La Laguna, E-38206 La Laguna, Tenerife, Spain\\
}

\date{Last updated ...}

\pubyear{2022}

\begin{document}
\label{firstpage}
\pagerange{\pageref{firstpage}--\pageref{lastpage}}
\maketitle

\begin{abstract}

\noindent The X-ray light curves of accreting black holes and neutron stars in binary systems show various types of quasi-periodic oscillations (QPOs), the origin of which is still debated. 
The Relativistic Precession Model identifies the QPO frequencies with fundamental time scales from General Relativity, and has been proposed as a possible explanation of certain types of such oscillations. Under specific conditions (i.e., the detection of a particular QPOs \textit{triplet}) such a model can be used to obtain self-consistent measurements of the mass and spin of the compact object. So far this has been possible only in the black hole binary GRO J1655-40.
In the {\it RXTE/PCA} data from the 1999-2000 outburst of the black hole transient XTE J1859+226 we found a QPO triplet, and used the the Relativistic Precession Model to obtain high-precision measurements of the black hole mass and spin -  $M$~=~(7.85$\pm$0.46)~M$\odot$,  $a_{*}$~=~0.149$\pm$0.005 - the former being consistent with the most recent dynamical mass determination from optical measurements.
Similarly to what has been already observed in other black hole systems, the frequencies of the QPOs and broad-band noise components match the general relativistic frequencies of particle motion close to the compact object predicted by the model. 
Our findings confirm previous results and further support the validity of the Relativistic Precession Model, which is the only electromagnetic-measurement-based method that so far has consistently yielded spins close to those from the gravitational waves produced by merging binary black holes.

\end{abstract}




\section{Introduction}

Quasi-periodic oscillations (QPOs) are common features in the X-ray emission of accreting compact objects, and are thought to arise in the innermost regions of the accretion flow. In a power density spectrum (PDS) they take the form of relatively narrow peaks, the centroid frequency of which can be associated with dynamical motion and/or accretion-related time-scales. QPOs have been known for several decades, although their nature remains elusive, and a plethora of models exists to describe their still debated origin \citep[see, e.g.,][for a recent review]{Ingram2019}.

The most common type of QPO observed in black hole (BH) X-ray binaries - the Type-C QPO \citep{Casella2005} - shows a varying centroid frequency ($\sim$0.1-30Hz), and has been often explained as the result of Lense-Thirring precession of matter around a compact object \citep[e..g.,][]{Stella1998, Ingram2009}. QPOs with centroid frequencies around a few hundred Hz (up to $\sim$500~Hz), although very rare, have been observed in a handful of BH systems \citep[e.g.,][]{Belloni2012}, either isolated or in pairs. In analogy with the case of neutron stars \citep[][]{vanderKlis1997}, such high-frequency narrow features have been named lower and upper high-frequency QPOs (HFQPOs).

The theory of General Relativity (GR) predicts that bound orbits of particles in a gravitational field are described by three  fundamental frequencies: the orbital frequency ($\nu_{\phi}$) and two epicyclic frequencies - vertical and radial ($\nu_{\theta}$ and $\nu_{\rm r}$, respectively). These frequencies combined yield two more frequencies used to describe the evolution of the orbits in time, i.e. the nodal precession frequency ($\nu_{\rm nod}$ = $\nu_{\phi}$ - $\nu_{\theta}$), and the periastron precession frequency ($\nu_{\rm per}$ = $\nu_{\phi}$ - $\nu_{\rm r}$). According to GR the motion of matter in the proximity of a BH should carry the signatures of strong-field gravity effects. Therefore, QPOs are naturally expected to be a likely products of relativistic effects, and hence, if properly understood, they represent both a powerful diagnostic of the physics of accretion and a GR test. 

The Relativistic Precession Model \citep[RPM][]{Stella1998,Stella1999,Stella1999a} associates the nodal, periastron precession, and orbital frequency with the type-C QPO, the lower HFQPO and upper HFQPOs, respectively, observed in the light curves of accreting BHs and neutron stars. \cite{Motta2014a, Motta2014b} showed that the RPM provides a natural interpretation for the QPOs and broad noise components observed in the PDS of the BH binaries GRO J1655-40 and XTE J1550-564. These authors also showed that the RPM provides an effective method to self-consistently estimate the mass and the spin of a BH based only on the detection of QPOs. 
Until recently, GRO J1655-40 was the only BH X-ray binary for which the three QPOs relevant to the RPM were detected simultaneously. In this paper we report on the detection of a QPO triplet in the PDS from the BH system XTE J1859+226, based on which we derived the BH mass and spin using the RPM. We then compare our mass estimate with the latest dynamical mass measurement reported in an accompanying paper by Yanes-Rizo et al. (submitted).

\section{Observations and data Analysis}\label{Sec:Obs}

XTE J1859+226 is a bright black-hole transient that went into outburst in 1999 and was extensively observed by the Rossi X-Ray Timing Explorer (RXTE), reaching a maximum flux of $\sim 5\times 10^{-8}$erg/cm$^2$/s \citep[][]{Zurita2002}. During the outburst, the source followed an evolution typical of many of these systems, going through all the main states and showing all three types of Low-Frequency QPOs \citep[see, e.g.,][]{Remillard2006, Casella2005, Motta2017}, including type-C QPOs. A full analysis of the QPOs in this system can be found in \cite{Casella2004}.

\begin{figure}
\centering
\includegraphics[width=0.43\textwidth]{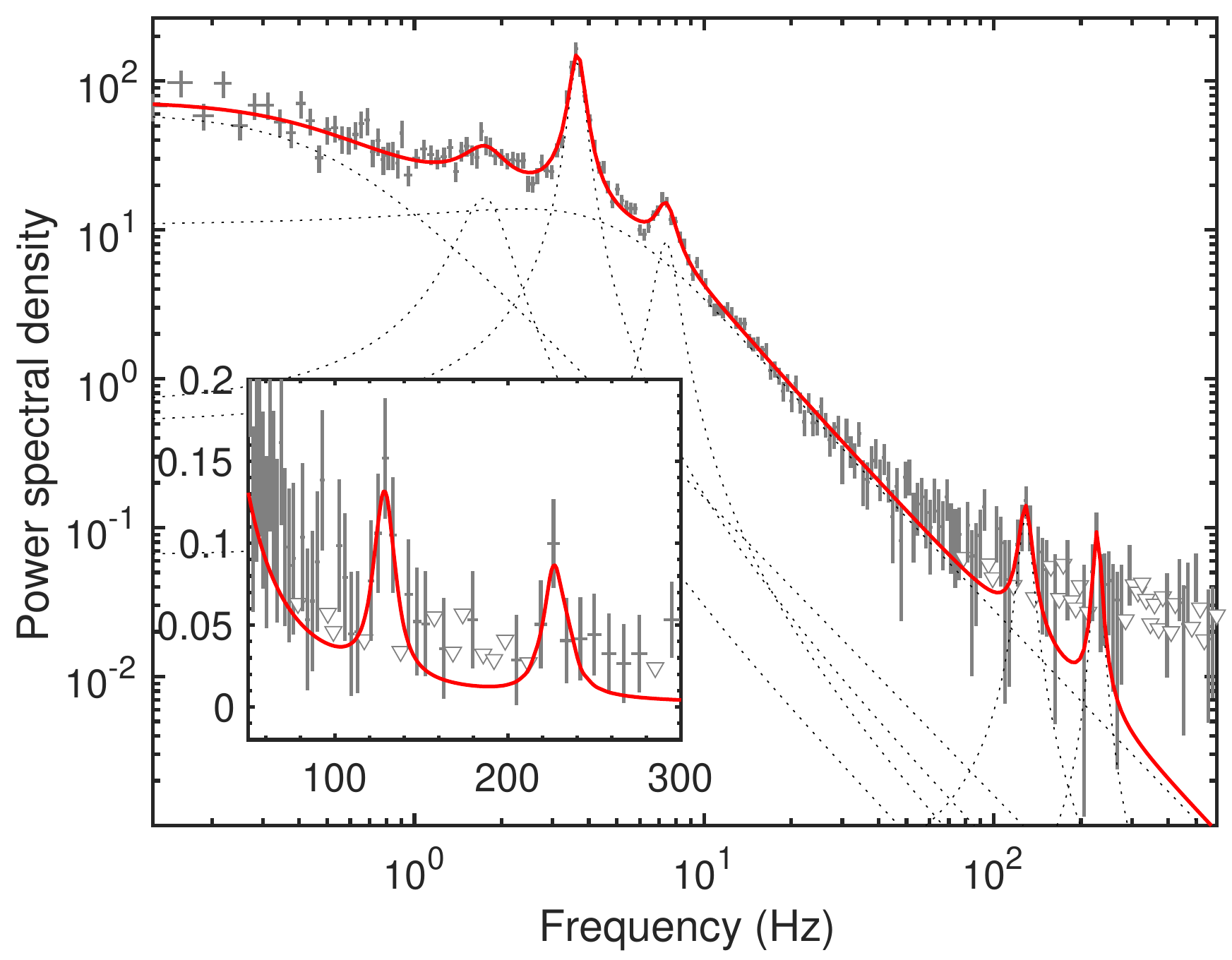}
\caption{Power Density Spectrum from Obs. 40124-01-07-00, including a QPO triplet. The red solid line marks the best fit model. Upper limits are shown as upside down triangles. All the best-fit model components are displayed. The inset shows a zoom of the HFQPO region.}
\label{fig:powsp} 
\end{figure}

We re-examined 130 archival RXTE observations of XTE J1859+226 taken with the Proportional Counter Array (PCA), collected between November 11th 1999 and July 24th 2000 (MJD 51462 - 51749). Such observations include a mix of \textsc{single-bit} mode data (covering the absolute channel range 0–35), and high-time-resolution \textsc{event} mode data (above absolute channel 36), which we combined to achieve coverage of the entire available energy range.
Following \cite{Motta2014a}, we considered each observation individually and computed average PDS using custom software under IDL  (GHATS\footnote{\url{http://www.brera.inaf.it/utenti/belloni/GHATS_Package/Home.html}}). 
We initially extracted PDS using various combinations of extraction parameters (interval length, energy range, time resolution), and we selected those that maximise the amplitude of the HFQPOs we found in the data (see below). 
We generated PDS using events collected in the absolute channel range 0-71 ($\approx$2-30 keV) employing 32~s long data segments, and a Nyquist frequency of 2048 Hz. We normalised the PDS from each data segment according to \cite{Leahy1983}, we averaged them obtaining an individual PDS per observation\footnote{We decided to extract one average PDS per observation as the exposure of the observations we considered is relatively constant ($\sim$1\,ks). }, and subtracted the contribution due to Poissonian noise \citep{Zhang1995}. 
Each PDS was fitted using the XSPEC package, adopting a one-to-one energy-frequency conversion and a unit diagonal response matrix. We fitted the noise components with a number of broad Lorentzian shapes, and QPOs and their harmonic peaks with one or more narrow Lorentzians \cite{Belloni2002}. We followed the method outlined in \cite{Motta2014a,Motta2014b} and we classified the narrow features identified in the PDS and selected only observations showing at least a significant type-C QPO. The relevant parameters from the best fit of the PDS are given in Tables \ref{tab:all_QPOs}. 
In the data from XTE J1859+226 we found two observations containing HFQPOs. The first,  40124-01-07-00, includes two HFQPOs and a Type-C QPO, i.e. a QPO triplet. The second, 40124-01-23-01, includes a HFQPO and a Type-C QPO, i.e. a QPO doublet. The PDS from observation 40124-01-07-00 is shown in Fig. \ref{fig:powsp}, while the best fit parameters for both the QPO triplet and doublet are reported in Tab. \ref{tab:triplets}.

\begin{figure}
\centering
\includegraphics[width=0.47\textwidth]{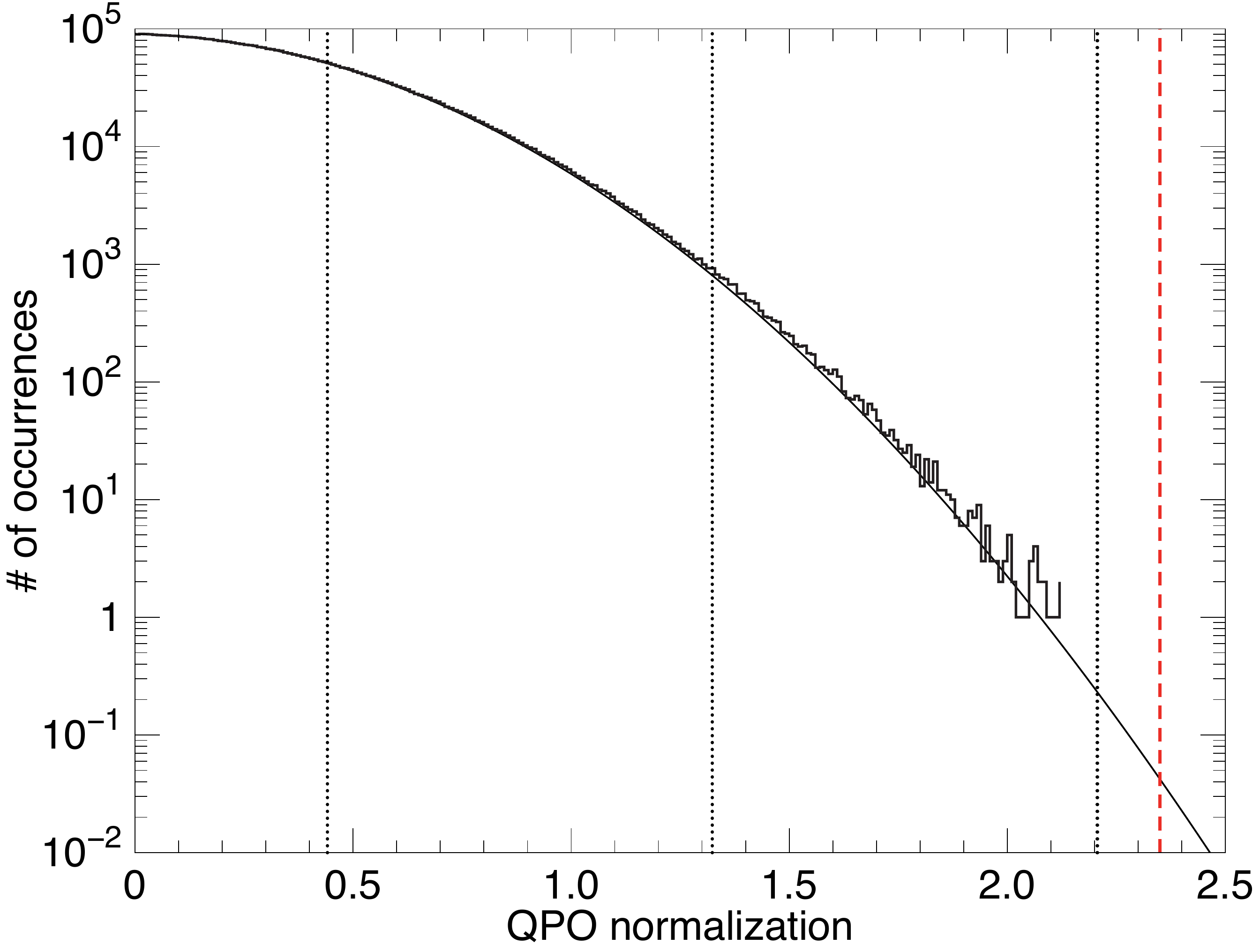}
\caption{Distribution of QPO normalisations obtained from the simulation described in the text for the lower HFQPO of the triplet. The black solid line is the best-fit Gaussian model to the distribution. The vertical red dashed line marks the value of the best-fit normalisation from the real data, and the vertical dotted lines indicate the values corresponding to 1,3,5 $\sigma$.}
\label{fig:pred_phi} 
\end{figure}

\subsection{HFQPOs significance estimate}

While an accurate estimate of the significance of type-C QPOs is usually unnecessary (being such strong features unmistakably significant when found in a PDS), the same is not true for narrow high-frequency signals, which tend to appear where the spectrum is dominated by red noise. The red noise, in many cases consisting of the high-frequency part of a band-limited noise component, is well-approximated by a simple power law, although it fluctuates wildly around the (unknown) true power spectrum. The above makes it difficult to distinguish real narrow quasi-periodic features from random fluctuations in the noise \citep[see, e.g.,][]{Vaughan2005,Vaughan2010}. 

In order to assess the chance probability of detecting the high-frequency peaks we performed the following simulations. For each HFQPO detection\footnote{Two HFQPOs are detected in the same observation, but since they are narrow peaks well separated in frequency we treated them independently.} we ran $10^7$ simulations of a power-law noise, with normalisation and slope drawn from a normal distribution about the best fit values, adding a constant Poissonian noise at the level of 2. From each model, we simulated a PDS with the same observational parameters of the real one (frequency binning $M$ and number of data intervals averaged $W$) following a $\chi^2$ distribution with $2MW$ degrees of freedom (d.o.f.), scaled by a factor $(1/MW)$  \citep[see][]{VdK1989}.
We fitted the resulting PDS with a power law plus a constant, then we divided the simulated PDS by the best fit model, multiplying by two to recover a $\chi^2$ distribution with $2MW$ d.o.f.. The final (flat) spectrum was fitted with a model consisting of a constant and a QPO with frequency fixed to the best fit value of the HFQPO under examination, and a $Q = \frac{\nu}{\Delta \nu}$ constrained between 6 and 20. Such values were chosen to guarantee that the fitted feature is narrow enough to match the HFQPOs observed in other sources \citep[see, e.g.,][]{Motta2014a}, and to avoid fitting with a feature too narrow to be realistic (i.e. FWHM less than a few bins).  This procedure led to a distribution of best-fit normalisation of QPOs detected from a model with \textit{no QPO}. As we did not constrain the QPO normalisation to be positive, half of the normalisation are negative. For clarity in Fig. \ref{fig:pred_phi} we only show the positive half of the distribution (which is symmetrical about zero) for the lower HFQPO of the QPO triplet. We fitted the distribution with a Gaussian and obtained from it the significance of the detections, reported in Tab. \ref{tab:triplets}. 

\begin{figure}
\centering
\includegraphics[width=0.47\textwidth]{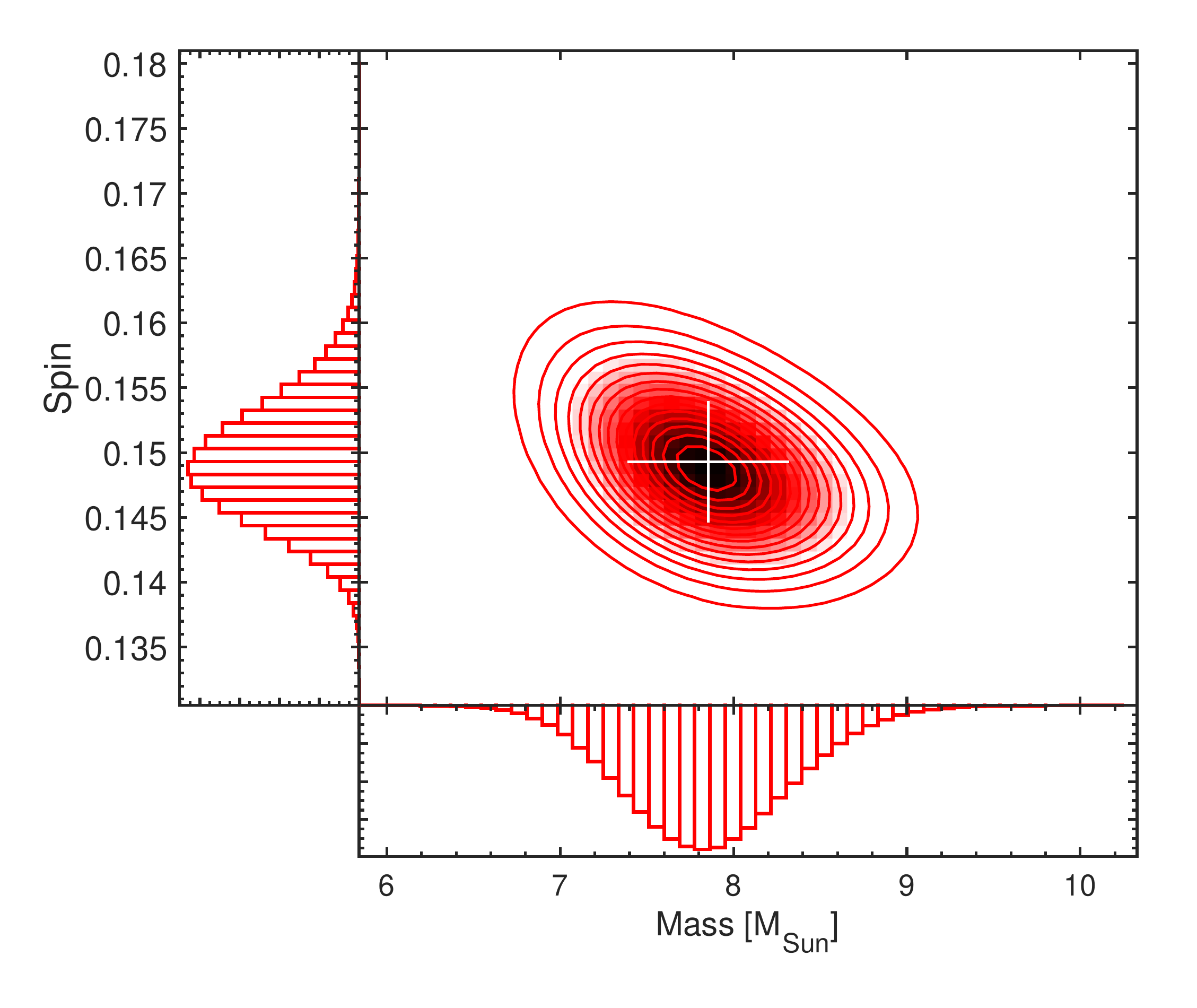}
\caption{Bi-variated mass-spin distribution, and marginal mass and spin distributions, resulting from a 10$^6$ steps Monte-Carlo simulation. The white cross in the main panel of the plot marks the central mass and spin value with 1~$\sigma$ uncertainties.}
\label{fig:spin_mass} 
\end{figure}

\begingroup
\setlength{\tabcolsep}{10pt} 
\renewcommand{\arraystretch}{1.2} 

\begin{table}
 \centering
\begin{tabular}{c c c}
\hline 
Obs. ID & QPO type   & Centroid frequency \\

\hline \hline
40124-01-04-00  &    Bump       &    61    $\pm$ 23$^\dagger$       \\
40124-01-04-00 	&	 C  	    &	 1.195 $\pm$ 0.006    \\
40124-01-05-00 	&	 C  	    &	 3.05  $\pm$ 0.01     \\
40124-01-06-00 	&	 C  	    &	 3.635 $\pm$ 0.008    \\
40124-01-08-00 	&	 C  	    &	 4.366 $\pm$ 0.007    \\
40124-01-09-00 	&	 C  	    &	 4.952 $\pm$ 0.007    \\
40124-01-10-00 	&	 C  	    &	 5.76  $\pm$ 0.02     \\
40124-01-11-00 	&	 C  	    &	 5.94  $\pm$ 0.01     \\
40124-01-18-00 	&	 C 	        &	 5.88  $\pm$ 0.01     \\
40124-01-19-00 	&	 C	        &	 5.185 $\pm$ 0.007    \\
40124-01-20-00 	&	 C 	        &	 6.43  $\pm$ 0.02     \\
40124-01-21-00 	&	 C 	        &	 6.33  $\pm$ 0.01     \\
40124-01-22-00G &	 C      	&	 6.06 $\pm$ 0.01      \\
40124-01-15-00 	&	 C 	        &	 6.15  $\pm$ 0.02     \\
40124-01-15-01G &	 C 	        &	 6.39 $\pm$ 0.03      \\
40124-01-15-02 	&	C	        &	 7.99  $\pm$ 0.2      \\
 40124-01-25-00 &	 C      	&	 6.18  $\pm$ 0.02   \\
 40124-01-14-00 &	 C*     	&	 8.85 $\pm$ 0.07    \\
 40124-01-16-00 &	 C*     	&	 7.71 $\pm$ 0.06    \\
 40124-01-17-00 &	 C*     	&	 7.47 $\pm$ 0.03    \\
 40124-01-23-00 &	 C*     	&	 6.94 $\pm$ 0.03    \\
 40124-01-15-02 &	 C*     	&	 7.9  $\pm$ 0.2     \\
 40124-01-26-00 &	 C*     	&	 7.27 $\pm$ 0.05    \\
 40124-01-28-00 &	 C*     	&	 7.83 $\pm$ 0.08    \\
 40124-01-28-01 &	 C*     	&	 7.54 $\pm$ 0.04    \\
 40124-01-29-00 &	 C*     	&	 7.62 $\pm$ 0.05    \\
 40124-01-31-00 &	 C* 	    &	 7.23 $\pm$ 0.03    \\
 40122-01-02-00 &	 C* 	    &	 7.26 $\pm$ 0.02    \\
 40124-01-58-01 &	 C* (?) 	&	 9.18 $\pm$ 0.07    \\
\hline 
\end{tabular}
\caption{Observation log including all the Type-C QPOs found in the archival data of XTE J1859+226. Observation ID, QPO type, and the QPO centroid frequency are listed. We mark with a * Type-C QPOs which are broader and fainter than canonical Type-C QPOs (see text for details). 
Note that the classification of the QPO detected in Observation 40124-01-58-01 is uncertain \citep[see][]{Casella2004}. Observations 40124-01-07-00 and  40124-01-23-01 are reported in Tab. \ref{tab:triplets} and are not included in this table.
$^\dagger$The frequency reported is the `bump' characteristic frequency, defined as $\nu_{\rm max}^2 = \nu^2 + \Delta/2$, where $\nu$ and $\Delta$ are the frequency and width, respectively, of the broad Lorentzian fitting the bump \citep[see][]{Belloni2002}. 
}\label{tab:all_QPOs}
\end{table}
\endgroup

\section{Results}\label{Sec:results}\label{sec:results} 

The results from our analysis are summarised in Tab. \ref{tab:triplets} and \ref{tab:all_QPOs}. We found one clear QPO triplet (see Fig. \ref{fig:powsp}), comprising a Type-C QPO and two HFQPOs (observation 40124-01-07-00), and a QPO doublet, consisting of a Type-C QPO and a HFQPO (Observation 40124-01-23-01). All the features detected are statistically significant: the upper and lower HFQPO in the triplet are detected at 4.25 and 3.45~$\sigma$, respectively, while the single HFQPO in the QPO doublet is detected at 4.04~$\sigma$. 

In the same data, we also found 29 QPOs which we classify as Type-C QPOs, all already reported in other works, although not necessarily classified as Type-C QPOs \citep[see, e.g.,][]{Casella2004}. Thirteen of these Type-C QPOs, which appear late in the outburst (i.e., close to the soft state), do not show the typical shape of a hard or hard-intermediate state QPO (i.e. high  amplitude and small FWHM). Instead, they appear as relatively broad peaks (quality factor Q = $\frac{\Delta\nu}{\nu} \sim$ 2) characterised by a lower rms amplitude. \cite{Casella2004} referred to such features as Type-C* QPOs.
As in \cite{Motta2014a,Motta2014b} we also searched for broad-band PDS components following the correlation originally found by \cite{Psaltis1999} (the PBK correlation), and successfully interpreted in the context of the RPM \citep{Stella1999}. We identified only one broad-band component that is detected together with a Type-C QPO (observation 40124-01-04-00), and which we tentatively classify as a $L_l$ broad component \citep[see][for details]{Motta2014a,Motta2014b}. According to \cite{Psaltis1999} and \cite{Belloni2002}, such a broad component is believed to be connected with the lower HFQPO. 

\bigskip

We used the centroid frequencies of the QPO triplet to estimate the mass and the dimensionless BH spin parameter ($a_* = J/M^2$, where $J$ is the BH angular momentum). To do so, we run a 10$^6$-steps Monte-Carlo simulation in which we adopted the analytical solution to the RPM equations given in \cite{IngramMotta2014} at every step. The above yields a BH mass M~=~(7.85$\pm$0.46)~M$\odot$,  a BH spin a$_*$~=~0.149$\pm$0.005, and a QPO emission radius $R_{\rm QPO}$~=~(6.85$\pm$0.18) $R_{\rm g}$ (where $R_{\rm g} = GM/c^2$ is a gravitational radius). 
From the spin we also derived an estimate of the innermost stable circular orbit radius $R_{\rm ISCO}$ = (5.50$\pm$0.016) $R_{\rm g}$ (see Tab. \ref{tab:BHpar}).
The result of the Monte-Carlo simulation is displayed in Fig. \ref{fig:spin_mass}, where we show the bi-variated mass-spin distribution, as well as the marginal mass and spin distributions (x and y axis, respectively). The cross in the main panel marks the central mass and spin value and the associated 1-sigma uncertainties. 

\begin{figure*}
\centering
\includegraphics[width=0.95\textwidth]{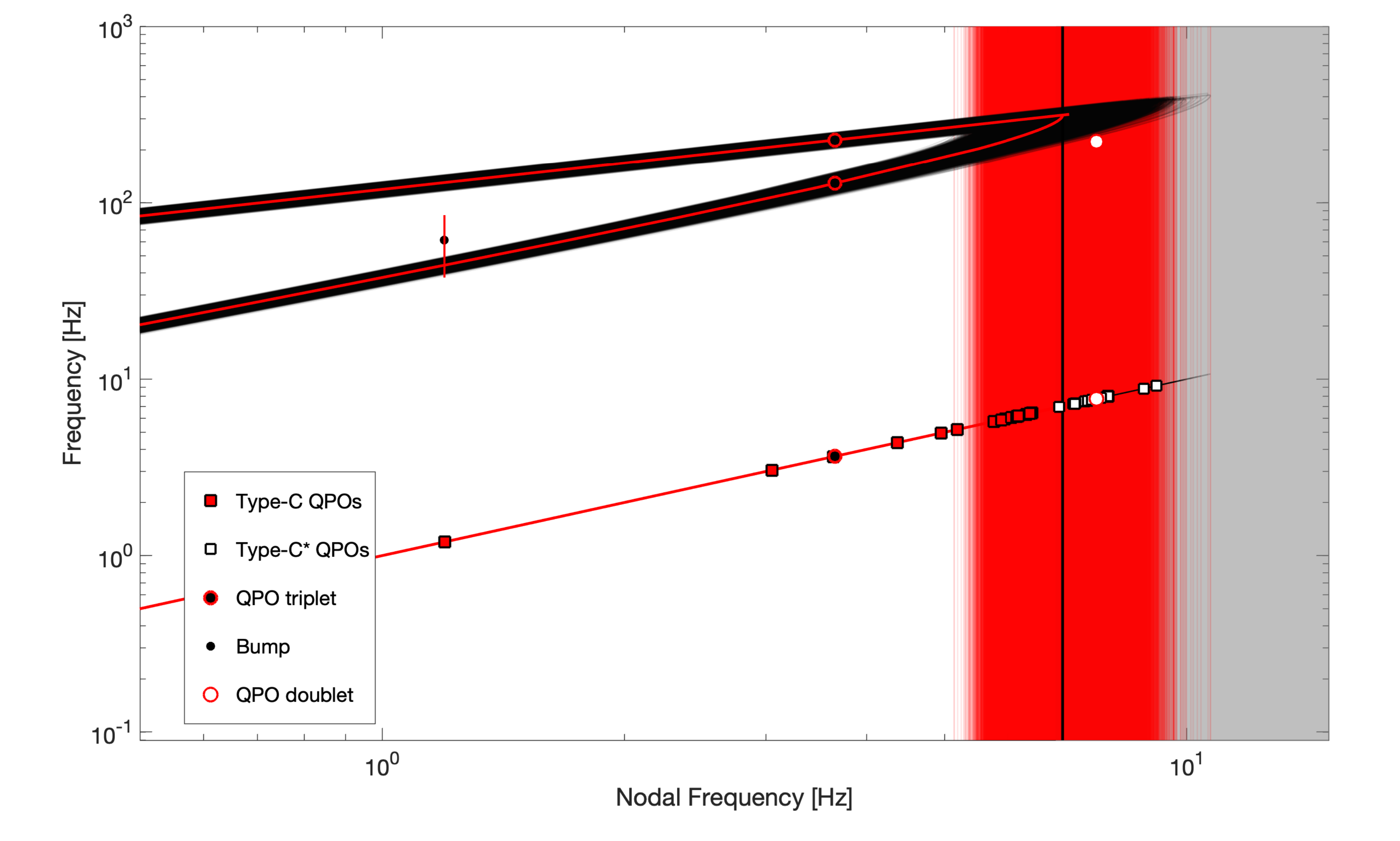}
\caption{Nodal precession frequencies (bottom line), periastron precession frequencies (middle lines), and orbital frequencies (top lines) plotted as a function of the nodal precession frequency, as predicted by the RPM using M~=~(7.85$\pm$0.46)~M$\odot$ and a$_*$~=~0.149$\pm$0.005. Note that we plotted \textit{all} the frequencies predicted for each mass-spin pair from the Monte-Carlo simulation in black, and marked in red the lines corresponding to the central values given above. The vertical red lines correspond to the predicted nodal frequency at $R_{\rm ISCO}$ for each mass-spin pair. In black the value corresponding to the central mass and spin values. The QPO triplet, the QPO doublet, all the Type-C QPOs, and the characteristic frequency of the broad PDS component (`bump') are indicated in the plot legend. Note that the type-C QPO frequencies are plotted against themselves to illustrate the frequency range spanned, hence the ‘correlation’ they form is an artefact.}
\label{fig:relativistic_frequencies} 
\end{figure*}

For each mass-spin pair obtained in the Monte-Carlo simulation, we predict the nodal, periastron precession, and orbital frequencies as a function of radius, as well as the value of $R_{\rm ISCO}$ and the corresponding nodal frequency. In Figure \ref{fig:relativistic_frequencies} we plot the three set of frequencies as a function of the nodal frequency, as well as the values of the nodal frequency at $R_{\rm ISCO}$ calculated in each step. The black bands in the plot are formed by the predicted frequencies for a varying radius predicted \textit{at each step} of the Montecarlo simulation (i.e. for each mass-spin pair), while the red thick lines indicate the frequencies predicted for the central values of mass and spin. Similarly, the vertical red band is formed by vertical lines marking the value of the nodal frequency expected at $R_{\rm ISCO}$ for each mass-spin pair, and the thick black vertical line corresponds to $\nu_{\rm nod}$|$_{\rm ISCO}$ derived from the central values of mass and spin. 

We marked on the graph the QPO triplet (red-edged black circles) all the Type-C QPOs detected in XTE J1859+226 (squares), the characteristic frequency\footnote{The characteristic frequency of a finite-width Lorentzian is defined as $\nu^{2}_{max} = \nu^2 + (\Delta/2)^2$, where $\nu$ is the central frequency and $\Delta$ is the width of the Lorentzian, see \cite{Belloni2002}.} of the broad PDS components (`bump' in Fig. \ref{fig:relativistic_frequencies}) versus the frequency of the Type-C QPO detected in the same observation (black dot), and the QPO doublet (red-edged white circle). 
All the type-C QPOs show characteristic frequencies that are consistent with being produced at either $R_{\rm ISCO}$ or at radii larger than $R_{\rm ISCO}$.  The lowest and highest frequency type-C QPOs we found are centred at 1.19 Hz and 9.1 Hz, respectively, which corresponds to a radius of $\approx$10 $R_{\rm g}$ and $\approx$ 5.1~$R_{\rm g}$. The HFQPO from the QPO doublet arises from a radius consistent with $R_{\rm ISCO}$ (where the two HFQPOs are predicted to have the same centroid frequency), although slightly lower than the central value given above, and at a frequency smaller than that predicted by the RPM.

QPOs observed in BH X-ray binaries typically show limited fractional widths, which provide clues on the geometrical structure of the region where QPOs are generated. As previously noted \citep[e.g.,][]{Stella1998}, the simplest assumption one can make is that QPOs arise in a narrow annulus in the accretion flow. Based on \cite{Motta2014a,Motta2014b} we estimated the radial size of such an annulus by apply a jitter $dr$ with variable magnitude to the emission radius at which the triplet QPOs are produced, and estimated the frequency range $d\nu_{i}$ (where i $\in$ \{nod, per, $\phi$\}) spanned by each predicted QPO. We started with a small jitter and increased it until $d\nu_{i}$ matched the full-width half maximum of the QPOs in the triplet. A jitter between 1.5 and 1.8 per cent around the estimated emission radius is enough to reproduce the observed width.

\begin{figure}
\centering
\includegraphics[width=0.45\textwidth]{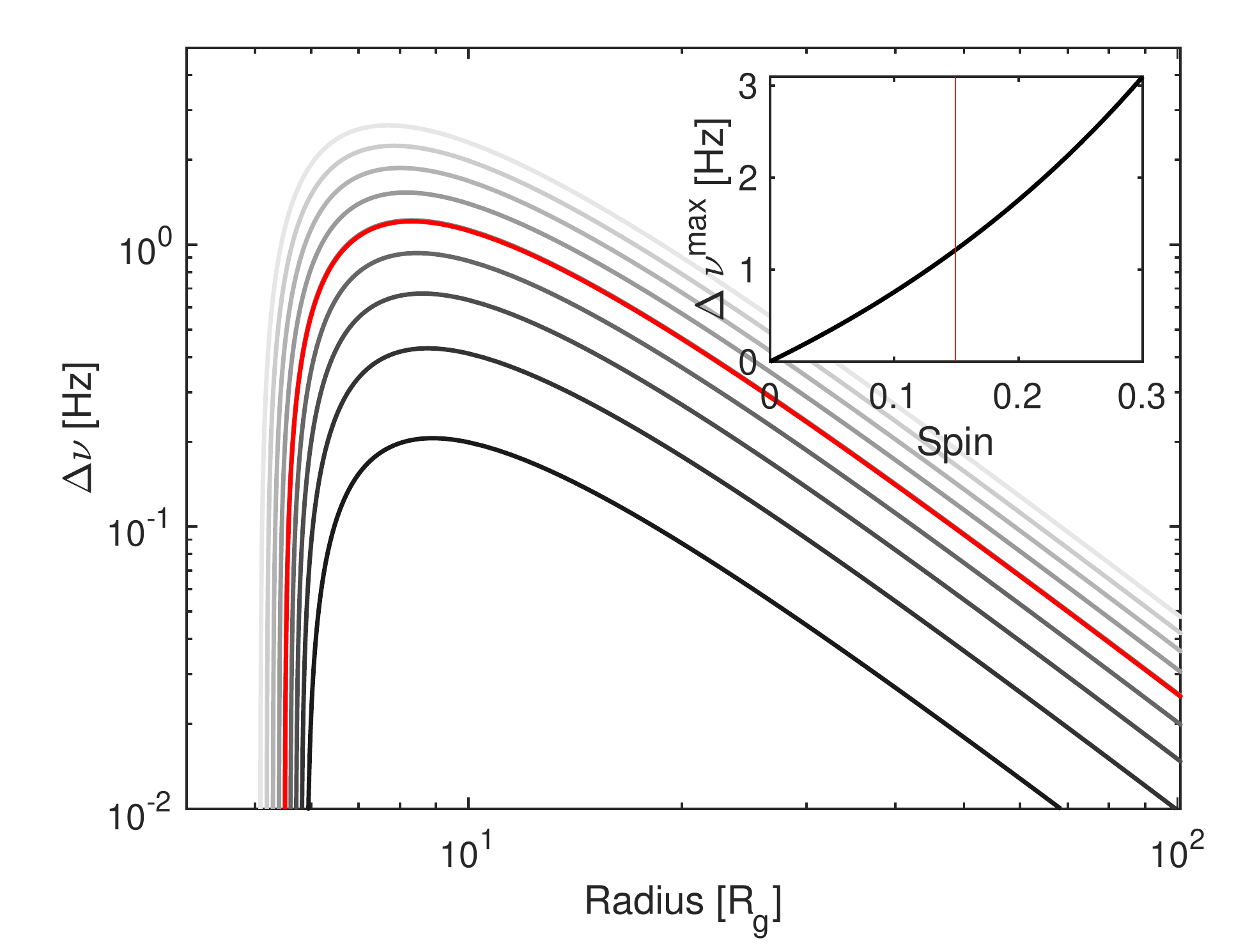}
\caption{Main figure: difference $\Delta \nu$ between the test particle nodal precession frequency and the global Lense-Thirring precession frequency, as a function of radius for various spin values. Lighter colours should correspond to higher spins. Inset: maximum difference between the test particle nodal and the global Lense-Thirring precession frequency and as a function of the spin. The spin value a$_*$ = 0.149 is marked in both plots. }
\label{fig:nuspread} 
\end{figure}

\begin{table}
\renewcommand{\arraystretch}{1.7}
 \centering
\begin{tabular}{c c c c c }
\hline 
 & QPO peak        &      Frequency [Hz]        &  Width [Hz]             &  Significance\\
\hline 
\hline 
\smallskip
\parbox[t]{2mm}{\multirow{3}{*}{\rotatebox[origin=c]{90}{Triplet}}} & Type-C         &      3.65$\pm$0.01            &  0.43$\pm$0.04            &    > 11.3~$\sigma$ \\
 & Lower HFQPO       &      128.6$_{-1.8}^{+1.6}$     &  12.5$_{-3.3}^{+5.9}$      &     4.25~$\sigma$  \\
 & Upper HFQPO       &      227.5$_{-2.4}^{+2.1}$     &  12.6$_{-4.1}^{+7.9}$      &     3.45~$\sigma$ \\
\hline 
\hline 
\parbox[t]{2mm}{\multirow{2}{*}{\rotatebox[origin=c]{90}{Doublet}}} & Type-C*     &  7.73$\pm$0.05               &  1.75$_{-0.03}^{+0.04}$    &    > 11.7~$\sigma$ \\
 & HFQPO         &  225.17$_{-3.2}^{+2.6}$      &  12.0$_{-1.0}^{+1.3}$      &     4.04~$\sigma$  \\
\hline 
\end{tabular}
\caption{A summary of the parameters of the narrow PDS features forming the QPO triplet (Observation 40124-01-07-00) and the QPO doublet (Observation 40124-01-23-01) described in the text. Note that the width of the upper HFQPOs in the triplet has been constrained to be $<$ 20~Hz. }\label{tab:triplets}
\end{table}

\begin{table}
\renewcommand{\arraystretch}{1.7}
 \centering
\begin{tabular}{c c c | c c c }
\hline 
\multicolumn{3}{c}{RPM parameters}  &  \multicolumn{3}{c}{Derived quantities}\\  
\hline
\hline
M        & = &      7.85$\pm$0.46 M$\odot$     & $R_{\rm ISCO}$               & = &      5.50$\pm$0.02 $R_{\rm g}$   \\
$a_{*}$                    & = &      0.149$\pm$0.005            & $\nu_{\rm nod}$|$_{\rm ISCO}$    & = &    7.05$\pm$0.60 Hz          \\
$R_{\rm QPO}$            & = &     6.85$\pm$0.18 $R_{\rm g}$      & $\nu_{\phi}$|$_{\rm ISCO}$    & = &   316.2$\pm$19.4 Hz          \\

\hline 

\end{tabular}
\caption{A summary of the BH parameters and associated quantities obtained via the RPM. Uncertainties are given at a 1-sigma level. }\label{tab:BHpar}
\end{table}

\section{Discussion}\label{sec:discussion}

The measurement of BH masses and spins is key to properly understand these relativistic objects, although it still represents a major issue in astrophysics. Accurate mass measurements can be performed  through challenging dynamical studies only in systems observed at a relatively high orbital inclination, and with a detectable companion \citep[see, e.g.,][]{Casares2014}. Alternatively,  when the companion star is not detected, indirect methods based on the properties of the H-$\alpha$ emission line from the quiescent accretion disc have been recently put forward to obtain the compact object mass \citep[see][and Casares et al. 2022, submitted]{Casares2015,casares2016,Casares2018}.  
Spins are typically inferred via spectroscopic studies relying upon the (often uncertain) identification of the inner disc truncation radius with $R_{\rm ISCO}$, which depends both on the BH mass and spin \citep[see, e.g.,][for a recent review]{Reynolds2021}. The Relativistic Precession Model provides an independent method to obtain precise and self consistent estimates of the mass and spin of a BH based solely on X-ray timing. When applied to high time-resolution data, such a method can yield extremely precise measurements. 

So far, a timing-based estimate of the mass and spin of a BH solely from a QPO triplet has only been possible for the BH binary GRO J1655-40 \citep{Motta2014a}. An estimate based on a QPO doublet (one HFQPO and a Type-C QPO) in combination with an independent mass measurement was performed for XTE J1550-40 \citep[][]{Motta2014b}. 
The RPM has also been successfully tested on neutron stars as well, where it has yielded promising results \citep{duBuisson2019}, even though a number of caveats had to be carefully considered since the simple Kerr metric used for BHs is not formally appropriate in the case of (non-magnetised) accreting neutron stars \citep[see discussion in, e.g.,][]{Maselli2020}.

In this work we examined 130 RXTE observations of the BH X-ray binary XTE J1859+226, collected during the 1999-2000 outburst of this system. We found a new QPO triplet - a Type-C QPO and two HFQPOs detected simultaneously - as well as a QPO doublet - a Type-C QPO and a HFQPO. We used the RPM with the centroid frequencies of the three simultaneous QPOs to obtain a precise estimate of the mass and spin of the BH, i.e.,  M~=~(7.85$\pm$0.46)~M$\odot$, $a_{*}$~=~0.149$\pm$0.005. 
If the RPM was an exact description of the behaviour of particles orbiting a BH, the uncertainties on the derived BH mass, spin and $R_{\rm QPO}$ would only come from the uncertainties on the measured QPO frequencies (which are dominated by the accuracy of the detecting instrument). However, the exact geometry of the emitting region might be a source of systematics. In Section \ref{sec:results} we assumed that the width of the QPO reflects the magnitude of the emission radius jitter. Based on the same assumption we can gauge the systematics related with the uncertainty on the radial extent of the region originating the QPOs. We run the same simulation described in Sec. \ref{sec:results}, but this time we assumed that the uncertainty on the QPOs centroid frequencies are equivalent to the FWHM of each QPO of the triplet. We obtained uncertainties a factor of 6 larger on the spin (relative error  $da \approx$ 16 per cent), and a factor of 2 larger on the mass (relative error $dM \approx$ 13 per cent) with respect to those reported in Tab. \ref{tab:BHpar}. Hence, we conclude that the systematics that could be affecting our mass and spin estimates are of the order $\sim$15 per cent.  

Our mass measurement is fully consistent with the value of the mass determined independently through the most recent spectro-photometric optical observations of XTE J1859+226 in quiescence collected with the 10.4-m Gran Telescopio Canarias and the 4.2-m William Herschell Telescope in 2017, i.e. M$_{\rm BH}$ = 7.8 $\pm$1.9 M$_{\odot}$ (Yanes-Rizo et al. submitted).
XTE J1859+226 is the second system for which a mass estimate has been possible both via dynamical measurement and through the RPM, and both in this and in the previous case (GRO J1655-40), the mass estimates from X-ray timing are remarkably consistent with the dynamical mass measurements, despite the fact that the masses of the two BHs are significantly different. This strengthens the validity of our method and essentially nullifies the possibility of a chance consistency of results.
%
Our spin measurement agrees with the only spin estimate we found in the literature, i.e. a $\in$ [0.1 - 0.4], reported in \cite{Steiner2013}, which is based on an empirical correlation between BH spins and jet power in accreting BH X-ray binaries, first identified by \cite{Narayan2012}. However, we note that such a correlation has not been confirmed in later works \citep[see][]{Russell2013b}, hence the validity of the aforementioned spin estimate remains uncertain.

Once the BH mass and spin are known, a number of predictions can be made. 
Assuming that the QPOs in the triplet arise from an annular region with mean radius equal to the emission radius $R_{\rm QPO}$, the radial extent of such an annular region should be approximately 0.1-0.2 $R_{\rm g}$, which is reproduced by a jitter of approximately 1.5 to 1.8 per cent around the QPO emission radius, consistently with what was found in GRO J1655-40, XTE J1550-564, and MAXI J1820+070 \citep[][]{Motta2014a, Motta2014b, Bhargava2021}.
Given  a mass and a spin, the RPM can also be used to predict the expected frequencies for the three QPOs relevant to the model as a function of radial distance from the BH. All the frequencies reach their highest value at the innermost stable circular orbit, where the radial frequency vanishes and the orbital and periastron precession frequencies coincide. The expected frequencies can then be compared with the observed QPO frequencies, and for each QPO a corresponding emission radius can be derived. All the QPOs we considered are consistent with being produced at a radius larger than $R_{\rm ISCO}$. However, around a nodal frequency of $\approx$7Hz ($\approx$5.5 $R_{\rm g}$), i.e. approximately in correspondence of the central value of ISCO, Type-C QPOs are replaced by Type-C* QPOs (broader and fainter than canonical Type-C QPOs). Type-C* QPOs are known to appear near the hard-intermediate to soft-intermediate state transition \citep[see, e.g.,][]{Casella2004, Belloni2016}, which has been speculated to happen when the accretion disc reaches the ISCO, and relativistic jets are launched \citep[see, e.g.,][]{Motta2019}. In this scenario, as already observed for the case of GRO J1655-40, part of the QPOs detected arise from radii slightly smaller than $R_{\rm ISCO}$. This is not particularly surprising  as the minimum allowed radius for the inner edge of a real accretion disc is expected to deviate from that predicted for a Shakura–Sunyaev disc \cite[see Discussion in][]{Motta2014a}. 

Perhaps surprising is the fact that the QPO doublet arises at a radius consistent with $R_{\rm ISCO}$, but the HFQPO appears at a frequency lower than what would be expected at that radius ($\approx$225 Hz instead of $\approx$316Hz). However, we note two things: first, the fact that only one HFQPO is detected supports the idea that the doublet arises from the ISCO, since here the radial frequency vanishes, and the periastron precession frequency equals the orbital frequency. Second, it is possible that additional forces counterbalance gravity near the ISCO, thus modifying the frequencies. For instance, magnetic fields are a necessary condition to generate jets in accreting systems, including relativistic jets that arise precisely across the hard-to-soft transition, hence when the accretion flow is expected to extend down to the ISCO \citep[see, e.g][]{Fender2009}. In the scenario of a magnetised accretion disc, the magnetic field could provide some additional support to the disc material against gravity, and this could affect the behaviour of matter near the ISCO. In the presence of a magnetic field with an azimuthal and vertical component (and no radial component\footnote{The configuration of the magnetic field inside the disc cannot have a significant radial profile as long as the disk is described in a good approximation by ideal MHD, because differential rotation would quickly drag the froze-in magnetic lines along the direction of the orbital motion and turn it into a toroidal magnetic field along the azimuthal direction.}) the orbital and radial frequencies of the accreting fluid (but not the vertical one) are expected to decrease proportionally to the magnitude of the magnetic field  (Pappas et al. in prep.). Therefore, the orbital and periastron precession frequencies are expected to increase more slowly near the ISCO with respect to the ideal, magnetic-field-free case, and to become equal at radii slightly smaller than that of the `ideal' ISCO. The magnitude of the change in the frequencies would depend on the magnetic field strength and its detailed configuration. 
Of course, the detection of a HFQPO at a frequency inconsistent with the predictions of the RPM might also indicate that the RPM is not the correct model for HFQPOs and thus other possibilities should be explored (see, e.g., the model proposed by \citealt{Fragile2016}). Tests involving other models are beyond the scope of this paper, and are left to a forthcoming work.

Differently from what observed in GRO J1655-40 and XTE J1550-564, the QPO triplet in XTE J1859+226 appears to be produced relatively far from ISCO, with a frequency 48 per cent smaller than the expected value at ISCO (in the case of  GRO J16655-40 and XTE J1550-564, the QPO triplet was produced at frequencies 30 per cent and 23 per cent smaller than the value at ISCO, respectively). 
This may be in apparent contrast with the implications of the PBK correlation, which suggests that the physical mechanism that produces HFQPOs near $R_{\rm ISCO}$ should give rise to broad-band PDS components at larger radii. In this sense, one might might expect to observe a two broad PDS components and a type-C QPO instead of a QPO triplet. However, in our data-set we do detect one broad-band PDS component simultaneous with a Type-C QPO at $\approx$1 Hz, which implies that the broad component arises from radii significantly larger than $R_{QPO}$ (the Type-C closest in frequency is observed at $\approx$3Hz). 
The fact that the QPO triplet arises relatively far from ISCO may also imply that the RPM might not be suitable to accurately model the motion of matter in the accretion flow, and global rigid Lense-Thirring precession of a radially extended accretion flow with outer radius $R_{\rm out}$ = $R_{\rm em}$ should be considered instead  \citep[see][]{Motta2018}. However, we note that at least for what concerns the nodal precession frequency, the difference $\Delta_{\nu}$ between the test-particle nodal precession frequency and the global rigid Lense-Thirring precession frequency depends strongly on the spin, and despite being non-monotonic and peaked at a radius close to $R_{ISCO}$, it decreases rapidly for decreasing spin (see Fig. \ref{fig:nuspread}). For a spin $a_{*}$ $\approx$ 0.149 and Mass $\approx$7.8, and for reasonable assumptions on the disc surface density profile (p = 0.6, see \citealt{Motta2018}), a difference of 1 Hz between the test particle and the global frequency corresponds to only an absolute variation in spin of 0.029. 

In this paper we presented the third case of a BH spin obtained via the RPM (the first two were presented in \citealt{Motta2014a} and \citealt{Motta2014b}). In all three cases the spin values we obtained are relatively low as compared to those typically obtained via X-ray spectral fitting techniques \citep{Reynolds2021}. Interestingly, we note that the timing-based estimates are instead close to the spin distribution emerging from an increasing number of gravitational wave events from the coalescence of compact objects binaries. On the one hand, our results stress once more that the inconsistency of the spin measurements obtained via any two (or more) of the available `electromagnetic' methods (timing, continuum fitting, reflection spectroscopy, and reverberation) is symptomatic of an inexact, or at least inaccurate, modelling of the data. On the other hand, \textit{if} the RPM is correct, our results alleviate the tension between the gravitational waves \citep[see][]{Wysocki2022,Golomb2022} and the electromagnetic spin measurements \citep[][]{Fishbach2022}, and supports the hypothesis that LIGO/Virgo BHs and X-ray binary BHs do not necessarily form two distinct populations \citep{Belczynski2021}.

\section{Summary and conclusions}

We presented the analysis of RXTE archival data on the BH X-ray binary XTE J1859+226, where we found three simultaneous QPO that can be associated with the nodal frequency, the periastron precession frequency, and the orbital frequency of particles in orbit around a spinning BH. Via the Relativistic Precession Model we obtained simultaneous and self-consistent measurements of the BH mass and dimensionless spin parameter, M~=~(7.85$\pm$0.46)~M$\odot$ and  $a_{*}$~=~0.149$\pm$0.005. The mass value is fully consistent with that obtained in the most recent optical
dynamical study (M$_{BH}$ = 7.8 $\pm$1.9 M$_{\odot}$), and the spin agrees with the only other estimate present in the literature. 
Beside providing a fully acceptable mass-spin solution for XTE J1859+226, the Relativistic Precession Model does not contradict the global Lense-Thirring precession interpretation of Type-C QPOs, of which it remains a limit case. 
The agreement between the predictions of the RPM and the timing properties of XTE J1859+226 further supports the hypothesis that certain types of QPOs commonly found in accreting BH X-ray binaries can be explained in terms of motion of matter in the close vicinity of a compact object.

This is the second time that a BH mass is obtained solely based on X-ray QPOs, and the third time a spin is derived via the RPM. The agreement between the BH masses from dynamical measurements and the timing values strengthens the validity of the RPM, while the fact that the timing-based spins are close to the emerging distribution of spins from gravitational waves supports the hypothesis that the merging binary BHs and the BHs in X-ray binaries might be part of the same population. 

\bigskip
\section*{Acknowledgements}

\noindent  All the authors thank the anonymous referee for providing useful comments which contributed to improving the quality of this paper. 

SEM and TMB acknowledge financial contribution from the agreement ASI-INAF n.2017-14-H.0, the INAF mainstream grant, and PRIN-INAF 2019 n.15.
S.M. thanks Andrea Maselli for having carefully read the manuscript. 
LS acknowledges financial contributions from ASI-INAF agreements 2017-14-H.O and  I/037/12/0; from “iPeska” research grant (P.I. Andrea Possenti) funded under the INAF call PRIN-SKA/CTA (resolution 70/2016),  and from PRIN-INAF 2019 no. 15.
TMD and JCV acknowledge support from the Consejer\'ia de Econom\'ia, Conocimiento y Empleo del Gobierno de Canarias and the European Regional Development Fund (ERDF) under grant ProID 2020010104 and the Spanish Ministry of Science under grants PID2020-120323GB-I00 and EUR2021-122010.
MAPT acknowledges support by the Spanish MINECO under grant AYA2017-83216-P and a Ram\'on y Cajal Fellowship RYC-2015-17854.

\section*{Data Availability}

\noindent Data from \textsc{RXTE} are publicly available in the NASA's HEASARC Data archive. The source code used is based on published work and commonly used analysis tools and techniques, available publicly in a number of repositories.



\bibliographystyle{mnras.bst}
\bibliography{biblio}

%





\label{lastpage}
\end{document}